To appear in Proceedings of SPIE Conference
*Astronomical Telescopes and Instrumentation 2010*

# Overview of *EXIST* mission science and implementation

J. Grindlay[*a], N. Gehrels[b], J. Bloom[c], P. Coppi[d], A. Soderberg[a], J. Hong[a], B. Allen[a], S. Barthelmy[b], G. Tagliaferri[e], H. Moseley[b], A. Kutyrev[b], G. Fabbiano[a], G. Fishman[f], B. Ramsey[f], R. Della Ceca[e], L. Natalucci[g] and P. Ubertini[g]

[a] Harvard Smithsonian Center for Astrophysics, 60 Garden St., Cambridge, MA 02138;
[b] NASA Goddard Space Flight Center, Greenbelt, MD 20771;
[c] Department of Astronomy, University of California at Berkeley, Berkeley, CA 94720;
[d] Department of Astronomy, Yale University, New Haven, CT 06511;
[e] INAF-Osservatorio Astronomico di Brera, Merate, Italy;
[f] NASA Marshall Space Flight Center, Huntsville, AL 35812;
[g] INAF – IASF, Rome, Italy

## ABSTRACT

The Energetic X-ray Imaging Survey Telescope (*EXIST*) is designed to i) use the birth of stellar mass black holes, as revealed by cosmic Gamma-Ray Bursts (GRBs), as probes of the very first stars and galaxies to exist in the Universe. Both their extreme luminosity (~$10^4$ times larger than the most luminous quasars) and their hard X-ray detectability over the full sky with wide-field imaging make them ideal "back-lights" to measure cosmic structure with X-ray, optical and near-IR (nIR) spectra over many sight lines to high redshift. The full-sky imaging detection and rapid followup narrow-field imaging and spectroscopy allow two additional primary science objectives: ii) novel surveys of supermassive black holes (SMBHs) accreting as very luminous but rare quasars, which can trace the birth and growth of the first SMBHs as well as quiescent SMBHs (non-accreting) which reveal their presence by X-ray flares from the tidal disruption of passing field stars; and iii) a multiwavelength Time Domain Astrophysics (TDA) survey to measure the temporal variability and physics of a wide range of objects, from birth to death of stars and from the thermal to non-thermal Universe. These science objectives are achieved with the telescopes and mission as proposed for *EXIST* described here.

**Keywords:** Black Holes; Gamma-ray Bursts; Pop III stars and galaxies; AGN and supermassive black hole census; coded aperture telescopes and imaging detectors; rapid optical/nIR imaging and spectroscopy at zodiacal limit levels.

## 1. INTRODUCTION

Black holes, with masses ranging from stellar to supermassive (~$10^{0.5\text{-}10}$ $M_\odot$), are now recognized as fundamental to the very formation of galaxies and, by extension, their constituent stars, planets and perhaps (even) life. Although it is only the supermassive black holes (SMBHs) that have masses correlated with the velocity dispersion of their host galaxy stars in their central quasi-spherical "bulge", pointing to feedback and self-regulated growth of both the SMBHs likely formed from accretion onto, and mergers of, what were originally stellar mass BH "seeds" that might have been as massive as the ~100 $M_\odot$ BH remnants of the first stars. Thus the study of black holes on all scales impacts a wide range of key questions in astrophysics and motivated the original and then later *EXIST* concepts[1,2] for a BH survey mission.

As the most compact massive objects in the Universe, with Schwarzschild radii $R_s = 2GM/c^2$ interior to which their mass M is cloaked, the gravitational potential energy available for release by matter with mass m falling into BHs ranges between $E_g \sim (0.06 – 0.42)mc^2$, for stationary vs. maximally rotating BHs, and so is by far the most efficient luminosity source known. When catastrophic core collapse of a rapidly rotating massive star occurs at the end of its nuclear burning lifetime, under certain conditions (only partly understood – see Woosley and Bloom[3] for a review) a Gamma-Ray Burst (GRB) occurs which releases for the several solar masses (1 $M_\odot$ = 2 x $10^{33}$ grams) rapidly accreted into the BH a total isotropic equivalent energy of $E_{iso} \sim 10^{52\text{-}53}$ ergs in hard X-rays and soft γ-rays with characteristic peak energies $E_{peak}$ ~0.3 MeV that are detected as a GRB. Given the characteristic (long) GRB timescales of $\tau_b$ ~10 s for the radiation to emerge

* josh@head.cfa.harvard.edu; phone 1 617 495 7204; fax 1 617 495 7356

(in a jet), the GRB luminosity $L_b \sim E_{iso} / \tau_b \sim 10^{51\text{-}52}$ erg/s in detectable radiation is some $\sim 10^4$ times larger than the most luminous "steady" beacons from the distant-early Universe, quasars, which are persistently accreting SMBHs. It is this incredibly bright luminosity which makes the GRBs the most powerful probes to illuminate the structure of the early Universe. Conveniently, a GRB at redshift z has its time variability stretched by a factor (1 + z), so that the power law decay of the X-ray, optical and radio "afterglow" emission is dilated by this factor, thus enabling a GRB at z = 9 to be observed 10X earlier in the burst decay than a GRB at z =0.1 (say) observed at the same fixed delay.

Within the past 2y, GRBs have superceded quasars as the most distant objects with measured spectroscopic redshifts (Fig. 1a). GRB080913 (z = 6.7)[4] and GRB090423 (z = 8.2)[5,6] both exceed the highest redshift quasar (z = 6.43) discovered[7] with the Sloan Digital Sky Survey. GRBs are thus the most distant as well as bright beacons with which to study the early Universe. This novel prospect, discussed in more detail in ξ2.1, is the primary science goal of *EXIST* and drives the overall mission design.

The study of SMBHs, both as accretion-powered active galactic nuclei (AGN), ranging from relatively nearby low luminosity Seyfert galaxies to distant extremely luminous quasars, and as non-accreting (dormant) SMBHs, drives the second primary science goal for *EXIST* and related objectives, as described in ξ2.2. Accreting SMBHs may be highly obscured by gas and dust in their host galaxies, or they may be dominated by powerful beamed jets. Both are poorly surveyed, and maximizing sensitivity to both requires wide-field hard X-ray imaging surveys. Dormant SMBHs can be detected by the tidal disruption flares (TDFs) they produce by occasionally shredding normal stars which pass too close.

The requirements to achieve the first two science goals enable the third: the study of transients of all classes, and the physics of the time-variable high energy sky. The wide-field imaging needed to discover GRBs and rare classes (e.g. luminous blazars) or phenomena (e.g. TDFs) associated with SMBHs, directly enable multiwavelength studies for the emerging field of Time Domain Astrophysics, as described in ξ2.3.

We first summarize the three principal science objectives for *EXIST* in section ξ2, with initial reference to the *EXIST* design (Fig. 6). A broad overview of the mission to achieve the science goals is given in section ξ3. We then follow with a detailed description of each of the three major telescopes and associated instruments proposed for the mission in ξ4, followed by the requirements for the spacecraft and mission plan in ξ5. In ξ6 we outline the proposed mission operations, and data analysis and the Guest Investigator (GI) program, followed by a brief Summary in ξ7.

## 2. PRIMARY SCIENCE OBJECTIVES FOR *EXIST*

### 2.1 Gamma-Ray Bursts as probes of the Early Universe

Again, GRBs (more specifically, "long GRBs", with durations ~2-1000s) have surpassed quasars and faint galaxies as the most distant objects for which redshifts have been measured by spectroscopy (cf. Fig. 1a). The highest redshift object so measured[5,6] is GRB090423 at z = 8.2 at which distance/time (by the standard concordant cosmology) the Universe was only ~630My old. Whereas recent *HST*/WFC3 observations[8] have given photometric redshifts (from Y and J band dropouts due to Lyman α absorption by neutral H in intervening galaxies) of objects which appear to be very faint (AB magnitude ~28.5) galaxies at z ~8-9, these are too faint for spectroscopy until observed with JWST after its launch in ≥2014.

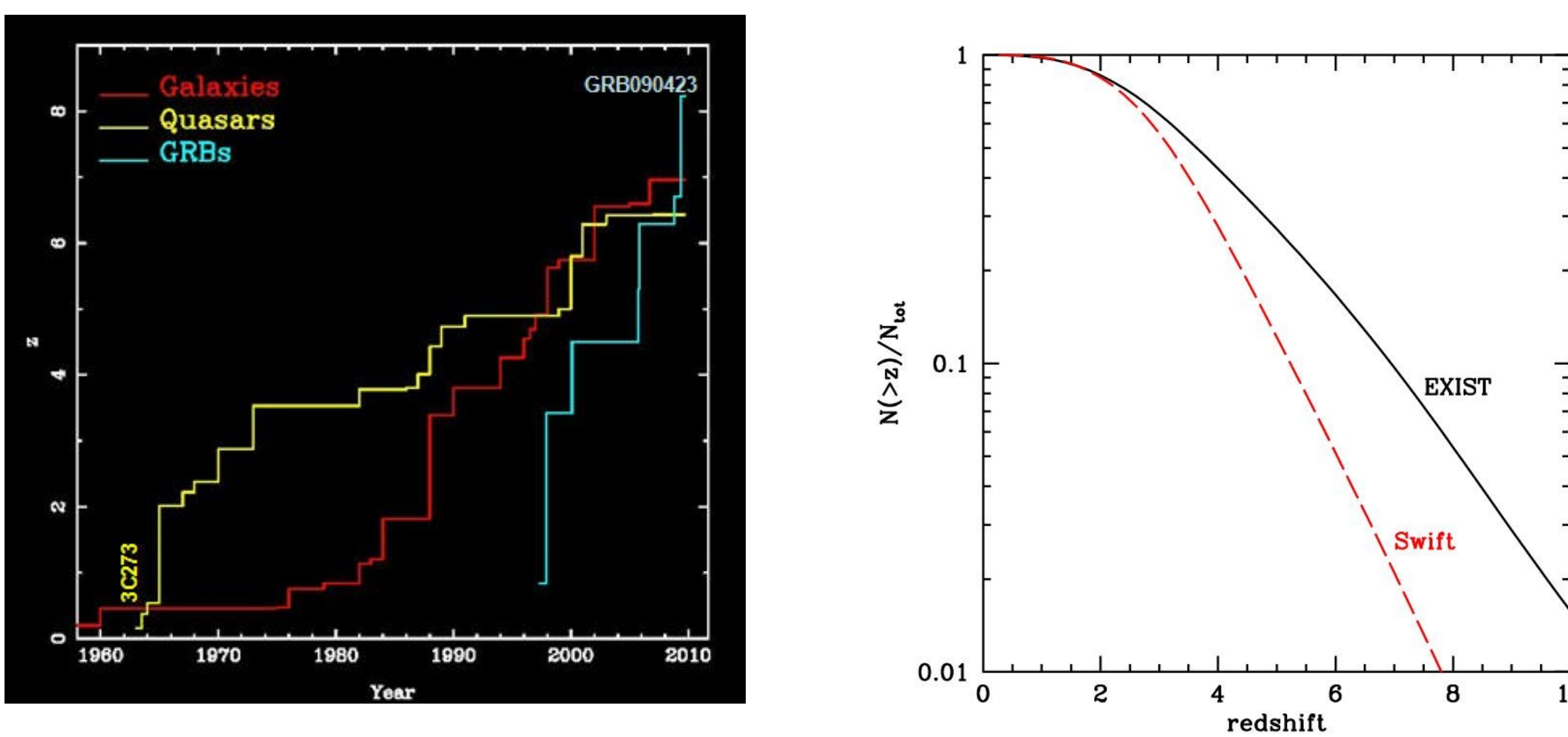

Figure 1. Left, a) Max z of GRBs, QSOs and galaxies vs. time; Right, b) expected fraction of GRBs vs. z for *EXIST* vs. *Swift*.

The increase after 2004 in the maximum z for GRBs was due to the launch of the *Swift* mission[9] with its more sensitive and wide-field Burst Alert Telescope (BAT) and rapid slewing to point a narrow-field X-ray Telescope (XRT) employing direct focusing to provide ~3arcsec X-ray afterglow positions to enable imaging identifications with the UV-optical telescope (UVOT), and to allow deeper searches for optical or near infra-red (nIR) counterparts. Over the 5.5y mission to date (July 2010), out of 526 GRBs detected and 441 XRT locations, 169 redshifts (all from ground-based telescopes) have been measured. The HET, SXI and IRT of the *EXIST* design (Fig. 6 and ξ3) each have ~10X the sensitivity and resolution (Table 1 and ξ4) so that virtually all GRBs will have redshifts measured on board. Fig. 1b shows the fraction of GRBs detected above a given z as predicted[10] for *Swift* vs. *EXIST* with the same GRB rate vs. star formation rate (SFR(z)). Normalizing to the actually measured 169 redshifts for *Swift* which are limited by ground-based telescope coverage and sensitivity, ~1.5 GRBs are expected above z = 8 from *Swift* which is consistent with that observed. For *EXIST*, with ~6X more GRBs expected above this redshift, and with most having their redshifts measured with a very sensitive optical-nIR telescope (IRT) *on board* (see ξ3), ≥10 and possibly 30 GRBs per year are expected to be measured at z >8. Over a 5y mission, this would enable the following use of GRBs as probes of the high-z Universe:

- *Measurement of SFR(z):* Long GRBs trace the massive star formation rate, which in turn is a shorter timescale measure of the total star formation rate, SFR(z). The sample of at least ~2000 GRBs with redshifts and ≥1000 with high resolution (R = 3000) spectra will provide the most comprehensive, and uniformly selected, measure of how star formation and thus galaxy formation and chemical evolution, have evolved over early cosmic time.

- *Measurement of Z(z):* Each GRB redshift spectrum will provide constraints on the metallicity, Z, of the host galaxy from absorption lines in its spectrum. GRBs bright enough (AB <20) to have high resolution (R =3000) spectra of their afterglows will provide direct measures of metal line strengths and thus well determined Z(z) values for perhaps ~500 GRBs vs. the limited sample now measured[11]. For the larger sample with AB >20-24, which will likely also include most with z >6, some can be measured at comparable resolution with JWST and future ~30m class telescopes on the ground (e.g. GMT/TMT/ELT).

- *Measurement of EoR(z):* The large sample of GRBs at z ~8-12 expected from *EXIST* will provide in situ constraints on the epoch of reionization (EoR) and its evolution over redshift. This is one of the key priorities for the mission and motivates the inclusion of the R = 3000 spectrograph for the IRT, which allows measurement of the shape of the red damping wing of the Lyman α absorption line to measure the ionization fraction in the local IGM vs. the host galaxy[12]. Simulated IRT spectra are shown in McQuinn et al[12] that demonstrate the feasibility, and Grindlay et al (2010) show that *Swift* GRB lightcurves redshifted to z = 12 indicate that a significant fraction (>50%) could be measured by *EXIST*. Since WMAP-7 CMB polarization data have obtained[13] an integrated constraint for the "midpoint" of the EoR to be at z = 10.5 ±1.2 for 50% ionization fraction, assuming a "flash" reionization model, the *EXIST* results will cover enough sightlines to constrain the spatial patchiness of the "local" (at high z) IGM and the ionization fraction vs. redshift.

- *Measurement of the Pop III era:* Finally, a deep high redshift sample of GRBs from *EXIST* could make the first detections of the gravitational collapse of Pop III stars, the primordial first generation of massive stars, which probably formed over a range of redshifts z ~15-30 and collapsed to form ~$10^{2-3}$ $M_\odot$ BHs which may produce particularly luminous and long GRBs[14]. Their bright afterglow spectra arising from their shock-heated surrounding ISM would be devoid of metals and plausibly marked by HeII λ1640 absorption or emission from the pre-GRB surrounding HII region. Likewise, the afterglow X-ray spectra will be dust free and thus without absorption features[15] though these would be redshifted to the XUV. Since even JWST will not directly detect Pop III stars or the ISM spectra of their host primordial "galaxies" (dominated by dark matter halos), GRBs are the best hope to directly detect the (demise of) the first stars or, if Pop III stars make pair instability supernovae without GRBs (cf. Woosley and Bloom[3]), the first generation of Pop II stars from the initially enriched ISM. These are "guaranteed" to produce detectable/traceable GRBs.

As discussed in ξ3 and 4, the proposed 3-telescope *EXIST* payload (HET, SXI and IRT) makes these ambitious studies possible out to z = 19 given the 2.3μm cutoff of the IRT which in turn is set by the passive cooling (to a modest -30C) of the telescope optics to reduce thermal backgrounds at ~2μm by a factor ~$10^3$.

### *2.2* Revealing obscured, dormant and the first (?) supermassive black holes to *EXIST*

Supermassive (>$10^6$ M$_\odot$) black holes are fundamental building blocks of galactic nuclei and the surrounding central bulges of galaxies, as evidenced by the apparent correlation that the Bulge mass, measured by the Bulge stellar velocity dispersion, σ, is ~$10^3$ times larger than the SMBH mass measured by its accretion source properties and variability – the "M-σ relation". A recent best fit relation for a range of galaxy types with dynamical mass measures by Gultekin et al[16] gives scatter of ~0.2dex and a best fit relation $M_{SMBH} = 10^{8.12 +/-0.11} M_\odot (\sigma/200km/s)^{4.24}$. Yet, how do these masses evolve and over what ranges of z ? And since SMBH mass must increase over <$10^9$ y to power luminous quasars seen as early as z = 6.4, and this requires either or both rapid accretion and galaxy (and bulge) mergers, then most SMBHs at z >6 are heavily obscured while most at z ~0.1 are expected to be gas starved and dormant. Clearly this picture is too simple since, for example, 3 out of 4 of the closest AGN are obscured and the peak of the luminous quasar distribution with redshift is at z ~2. These are major questions that drive the science of a number of upcoming (JWST) and proposed (IXO) missions. But despite their power, their sample sizes are necessarily limited, as is their bandwidth. What is needed is a broad survey to measure SMBH demographics with minimal biases for obscuration or dormancy. As described in ξ3, this is where *EXIST* brings unique assets to bear on fundamental SMBH questions:

1. *Full sky coverage* for the survey, to enable discovery and measures of intrinsically rare objects (e.g. the highest luminosity objects, and jet-dominated systems, particularly Blazars at high z) or low duty cycle events such as luminous tidal disruption flares (TDFs) expected[17] when main sequence stars encounter quiescent SMBHs.
2. *Multiwavelength imaging and spectroscopy*, from hard X-rays (up to 600 keV) to nIR with both extremes being optimum for minimizing absorption while at the same time maximizing the ability to distinguish star formation from accretion power. These are key requirements for attempts to chart SMBH growth over cosmic time.
3. *Repeated, or at least multiple, observations*, and simultaneously across multiple wavelengths, to measure spectral energy distributions (SEDs) without the confusion of non-simultaneous coverage *and* to measure spectral variability (particularly in the ~5-50 keV band) for the direct constraints on emission region size, and thus SMBH mass that these can provide. As discussed below (briefly), repeated measures of flux and time variability allow both SMBH (or BH) masses to be measured directly (given broad band spectral coverage) and for otherwise undetectable SMBHs in quiescence to be discovered by their occasional TDFs.

The scanning all sky survey proposed for the first 2y of the *EXIST* mission and which then triggers a 3y followup pointing program (see ξ3), together with the sensitivities and configuration of the instruments, have been designed to

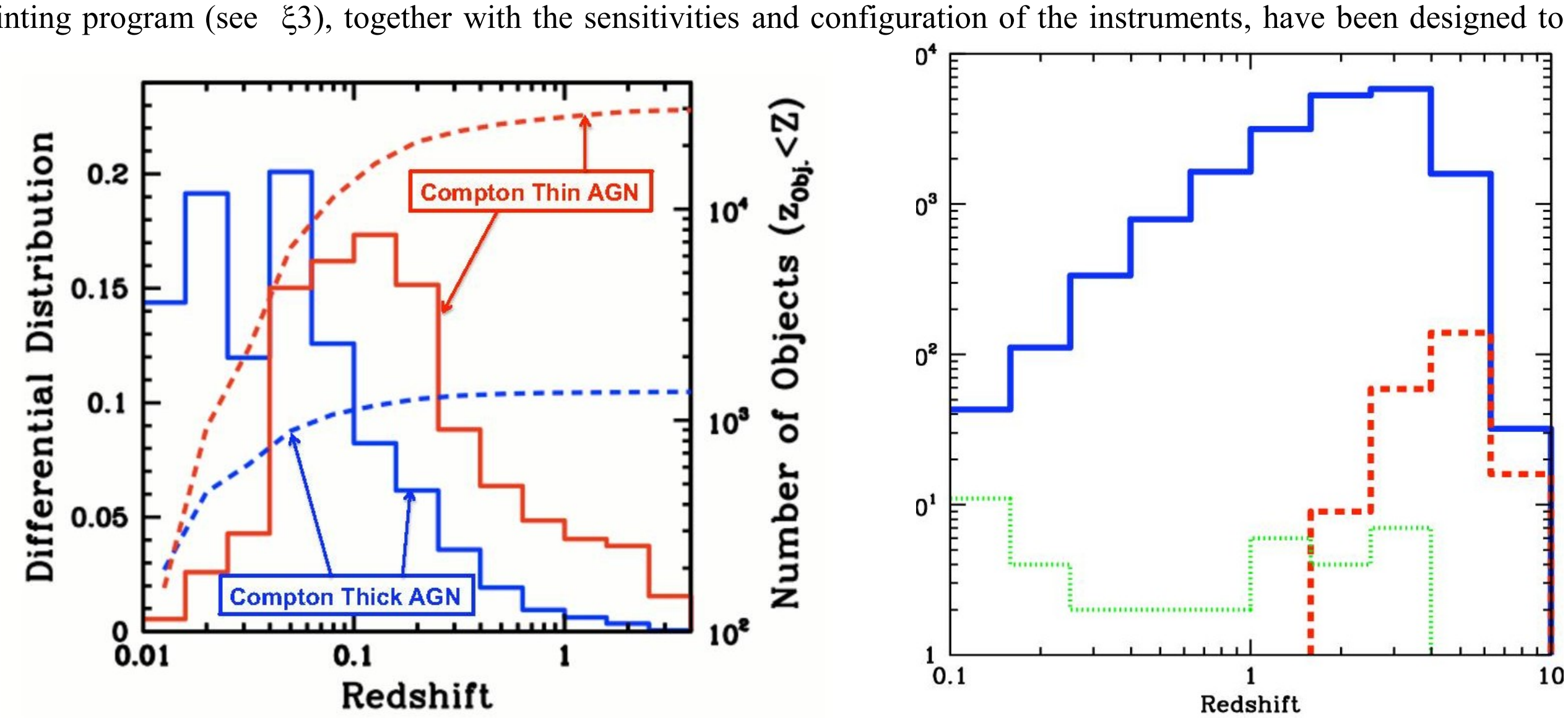

Figure 2. Left, a) Expected number of AGN of different classes (Compton thin vs. thick) vs. z for the 2y scanning full sky survey. Right, b) expected distributions of Blazars (see text; from Della Ceca et al[18]; DC09)

measure SMBH demographics which will constrain their origin and evolution.

*Obscured SMBHs:* The yields of unobscured (Compton thin) vs. obscured (Compton thick) AGN expected from the scanning 2y survey with flux limit $F_x$(10-40 keV) ≥8 x $10^{-13}$ erg $cm^{-2}$ $s^{-1}$ are shown in Fig. 2a (from DC09). Their differential (solid curves) vs. cumulative (dashed) distributions are plotted vs. redshift, using X-ray luminosity functions for each as described in DC09. The total sample size of ~28000 Compton thin, ~1400 Compton thick, and – most notably – some 19000 blazars is ~2 orders of magnitude larger than the current totals at hard X-ray energies (>10 keV) derived from any hard X-ray mission, most notably *Swift*/BAT or *INTEGRAL*/IBIS. The small field of view of future focusing hard X-ray missions (e.g. *NuSTAR*) will give smaller samples of Compton thick AGN despite their significantly greater sensitivity. The distribution of Compton thick objects, with absorption column densities NH ≥$10^{24}$ $cm^{-2}$, is particularly uncertain, being based primarily on the relatively small sample size from the XMM HBSS used (DC09) to scale to *EXIST* sensitivities and energy bands. Although Compton thick objects are by definition detected in their continuum spectra only above their ≥5 keV low energy cutoff energies (often accompanied by strong 6.4 keV fluorescent Fe line emission from surrounding gas), their "unabsorbed" high energy spectra are still reduced by Compton scattering. Thus for any flux limited survey, their apparently lower luminosity actually detected gives them a lower redshift distribution than the Compton thin sample. The most recent constraints on obscured (Type 2) AGN, with NH ~ $10^{22}$ $cm^{-2}$, but not Compton thick, is from the *XMM*-Cosmos wide field survey[19]. The fraction of these objects is much higher, reaching 90% at low luminosities ($L_x$ ~$10^{42}$ erg $s^{-1}$), but falling to ~20% at luminosities 3 orders of magnitude larger. The *EXIST* survey would provide the best constraints on the relatively rare high luminosity obscured AGN with its accurate measure of NH (from the 0.3 – 10 keV spectrum measured by the SXI – see ξ4.2) and their optical vs. nIR band colors (R – K) measured by the IRT (see ξ4.3). A large sample of objects like the luminous Type 2 QSO "XID 2028" discovered by Brusa et al.[19] would be found in the *EXIST* survey.

*Blazar probes of first SMBHS:* The redshift distribution expected for blazars (AGN with non-thermal beamed emission in jets) is shown in Fig. 2b and is the most dramatic promise of the SMBH survey expected for *EXIST*. This plot, also from DC09, is taken from Fig. 15 of Ghisellini et al.[20] (G10) and based on the luminosity function for blazars derived by Ajello et al.[21] from the 38 blazars (green dotted curve) detected at 15-55 keV in the BAT all-sky survey. The red dashed curve is the contribution from blazars with $L_x$ > 2 x $10^{47}$ erg $s^{-1}$ for their rest frame 15-55 keV luminosities, the luminosity limit of the 10 most luminous BAT blazars all at z >2 which were considered in detail by G10. All of these are Flat Spectrum Radio Quasars (FSRQs) with their inverse Compton peaks in the hard X-ray/soft γ-ray band (and so generally not detectable by *Fermi*/LAT), and for sub-Eddington accretion, their extreme luminosities imply that all must have SMBH masses >$10^9$ $M_\odot$. Thus, the very large Blazar sample expected will likely include extreme objects at z >6 that are *sensitive probes of the epoch of growth of SMBHs*. As shown by G10, the BAT blazar 2149-306 at z = 2.35 could be detected in the 2y *EXIST* scanning survey if it were at z =8, at which the short growth time might require major mergers and/or super-Eddington accretion. These objects might never be found in narrow-field surveys but could be then identified in the pointing phase of the *EXIST* survey by their continuum power law spectra from which the IRT would measure the redshift directly by the Lyman break in their redshifted spectra.

*SMBH mass measures:* Dynamical mass measures (not limits) for SMBHs in galactic nuclei are only available for 51 systems (Guletkin et al[16]), yet anchor the M-σ relation which in turn drives the discussion of black hole feedback in galaxies vs. SMBH growth. As noted in Guletkin et al, the dispersion in the relation is particularly large in spiral galaxies, and various sources of systematic error remain problematic. It is thus important to use independent techniques to measure or constrain SMBH masses, particularly for the luminous accreting systems (either obscured or not). X-ray timing offers just such a method, since accreting BHs show strong correlations between their variability timescales and BH mass when scaled for accretion rate (which determines the inner disk radius and thus minimum timescale) as given by the bolometric luminosity. McHardy et al.[22] showed that a sample of 10 X-ray bright Seyfert galaxies with variability timescales $T_B$ measured by the breaks in their power spectral density (PSD) as well as bright galactic BH X-ray binaries (Cyg X-1 and GRS1915+105) all fell on a relation $T_B \propto M_{BH}^{2.1}/L_{bol}^{0.98}$ for AGN (and BH-XRBs) with BH mass $M_{BH}$ and accretion luminosity $L_{bol}$. This relation is as expected from the scaling that minimum variability timescales must increase with BH size ($\propto M_{BH}$) and inversely with accretion radius which scales as Eddington accretion rate ($\propto L_{bol}/M_{BH}$). The 2y scanning survey will cover the full sky every 3h and measure the variability of all objects on timescales $\tau$ over which significant (>5$\sigma$) detections are made, which for the background limited wide-field HET are $\tau \propto 1/S^{0.5}$, where S is the source flux in some band. Since the logN-logS counts for AGN at the *EXIST* survey flux levels has slope -3/2, the number of AGN measureable should scale as $N(>T_B) \propto T_B^{3/4}$ as is approximately evident in Fig. 3 for AGN of the types as noted. To measure the PSD and a break timescale $T_B$ requires that lightcurves contain at least ~300 points (ideally >$10^3$) and thus range of timescales. Thus from Fig. 3, $T_B$ measures or limits, and thus values for $M_{SMBH}$, could be derived from a sample of ~400 AGN (a ~20X larger sample than now available from RXTE) which would include ~20 Compton thick objects and perhaps 30 blazars. Note that McHardy[23] finds that for 3C273 (a bright blazar), the $T_B$ ~1d is similar to much lower luminosity and SMBH mass Seyferts (due to the large $L_{bol}$ for 3C273 in the scaling relation), implying that time dilation from their relativistic jets does not affect the variability so that the variations originate in the (inner) disk, not beamed jet. This somewhat surprising result would be very well measured by *EXIST*, and compared with ~10 other brightest blazars that could be measured down to ~12h timescales to isolate the jet vs. disk variability region as well as the SMBH mass scaling relation for beamed objects.

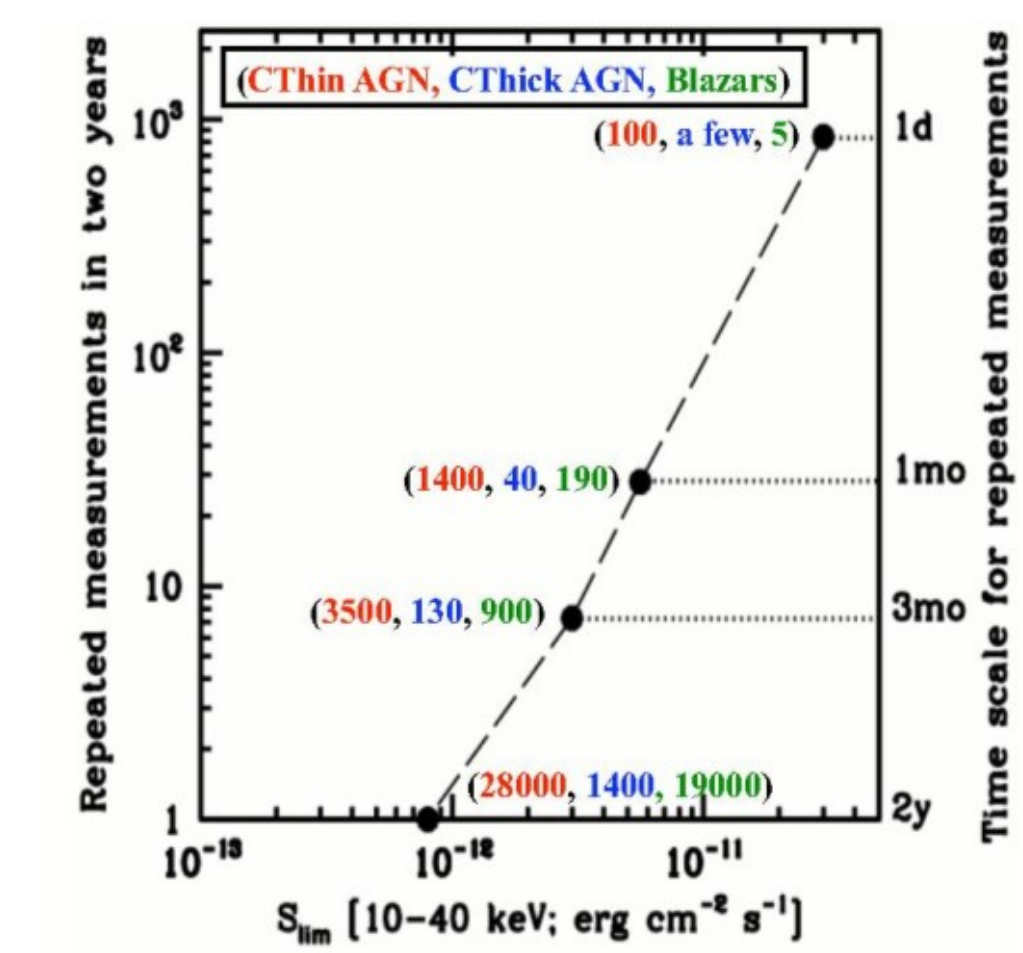

Figure 3. AGN numbers vs. minimum timescale observed in 2y survey (from DC09).

Two other points should be noted regarding the *EXIST* surveys and studies of SMBHs. First, the timing studies just outlined will continue during the 3y pointed mission phase since the ~2sr field of view (FoV) of the HET will include (for any pointing) ~1/6 of the AGN sample. Thus PSD measures are better (with longer baselines and more samples) than from the scanning survey alone, though scanning provides uniform sampling (full sky every 3h). Second, since the pointed mission phase is to identify and study sources from the scanning survey (see ξ3), it will by design include repeated pointed observations on a representative fraction (or all) of the ~400 bright AGN well covered in the scanning survey for more detailed timing and SMBH mass measures. These same observations will, of course, include both soft X-ray (0.3-10 keV) and optical-nIR spectroscopy, so that "definitive" broad band spectral variability and measures of $L_{bol}$ and other key indicators (e.g. emission line widths and reverberation delays) of $M_{BH}$ can be simultaneously obtained.

### 2.3 Time Domain Surveys

A priority goal for *EXIST* is to open new windows for Time Domain Astronomy (TDA) surveys and detailed temporal studies. The same repeated observations with the 2y scanning and then 3y pointing programs enable a wide varieity of transients and variables to be discovered and studied as shown in Fig. 4 (from Soderberg et al.[24]).

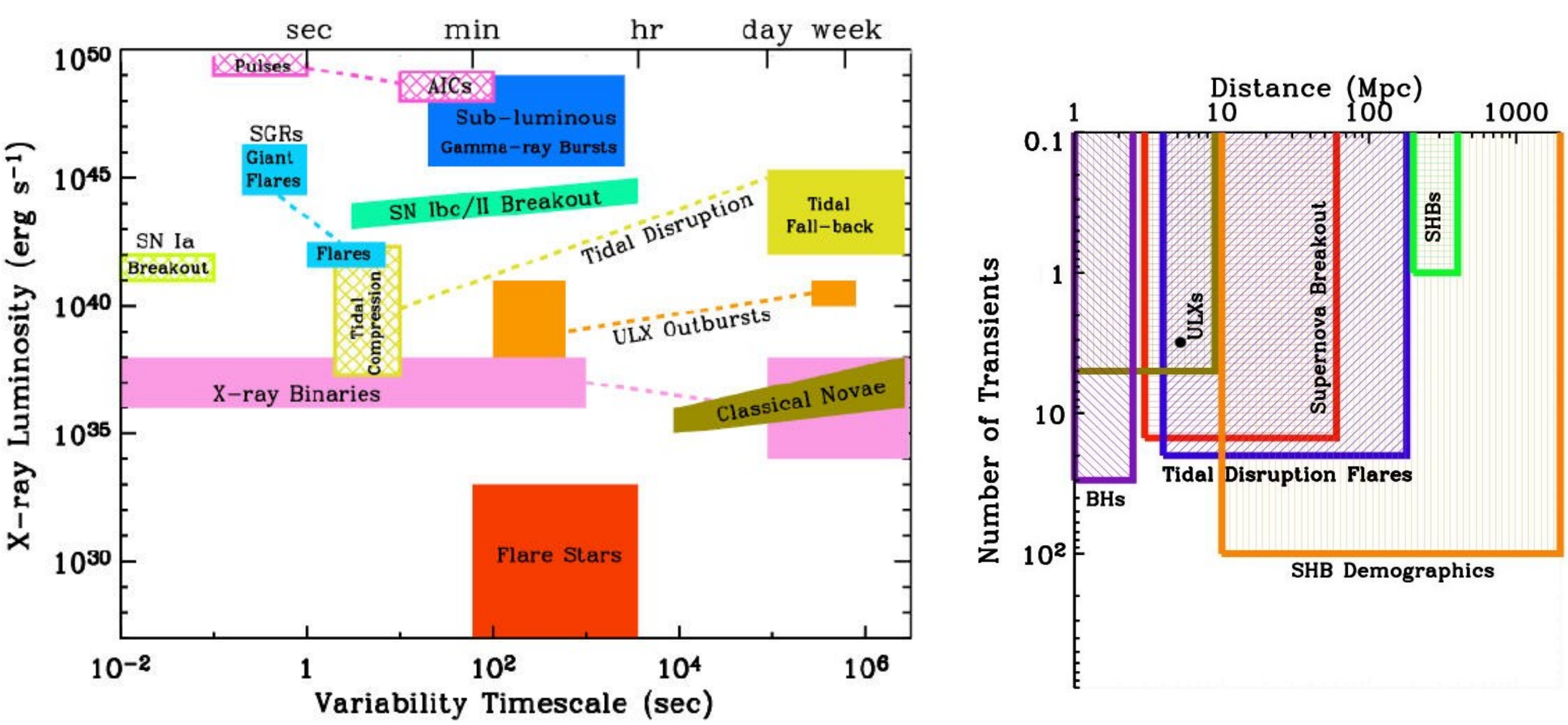

Figure 4. Left, a) X-ray transients to be studied in the *EXIST* temporal survey (Soderberg et al.[24]), and Right, b) predicted samples.

Each of these classes of objects can be explored in detail, without interference to the others, during the course of the scanning survey. We first mention a few with direct connections to the GRB and SMBH science priorities (ξ2.1, 2.2):

- *Sub-luminous GRBs vs. SNIbc/II breakouts:* Why did the first recognized GRB-SNIbc association, GRB980425, produce a low luminosity but detectable (with *BeppoSAX*) GRB whereas the first shock-breakout event, SN2008d, produce no triggerable GRB? How many of each class are there and how would the much larger sample expected from *EXIST* constrain the physics of GRB jets? Understanding this from a "local" sample would constrain models for how or under what conditions GRB jets could punch through much more massive and larger radius Pop III stars. The much greater sensitivity and bandwidth of the HET (vs. BAT) will produce a large sample of relatively nearby sub-luminous GRBs as well as a number of SN-breakout shock events, as summarized by Soderberg et al.[24].
- *Short GRBs vs. AIC vs. SGRs ?* Short GRBs (SGRBs), with observed durations (T90) <2sec and in fact usually <0.5sec, are less luminous than the long GRBs discussed in ξ2.1. The leading model for their production is the merger of two neutron stars, or possibly (in some cases) a NS-BH merger. The NS-NS merger model can be directly tested by the coincident detection of the SGRB with the gravitational wave signal (chirp) produced during the final inspiral and that for some can be detected with Advanced LIGO (>2014) out to large enough distances (~200 Mpc) that current estimates for total SGRB rates will produce several events per year. Depending on the (still) uncertain beaming of SGRBs, a fraction of these would be detected and precisely located by *EXIST* as SGRBs, thereby allowing a precision (<2%) measure of the Hubble constant *for each such GRB*[25]. The "no host" problem for SGRBs, where many are offset from their (plausible) host galaxies[26], is consistent with the model whereby NS-NS mergers are efficiently produced by dynamical interactions in the dense cores of globular clusters[27], which could be tested by the precise (<0.1") IRT locations for SGRBs detected with *EXIST* followed by still higher spatial resolution and deeper sensitivity imaging with JWST[28]. These same observations (*EXIST-ALIGO-JWST*) would at the same time rigorously test the two alternative SGRB production mechanisms, which though less likely may still account for some fraction of SGRBs: i) accretion induced collapse (AIC) of a magnetic white dwarf that could produce a SGRB like flash[29] but now without the gravitational wave chirp due to binary inspiral; or ii) a giant flare of a Soft Gamma-Ray Repeater (SGR), which have similar timescales and spectra to SGRBs but would arise from magnetar giant flares in disks of star-forming galaxies[30].
- *Tidal disruption flares:* The detections of TDFs due to the disruption of a (usually) main sequence star by a non-accreting (and so dormant) SMBH continue to mount along with models (Gezari et al[31]). The soft-X-ray luminous flares that marked their original discovery with *ROSAT* would be (generally) detectable above the 5 keV threshold for HET. However, a TDF but becomes much more detectable if the shock induced by the disruption event produces (as usually the case) a hard X-ray power law spectrum. In this case, *EXIST* should detect ~10-30 $y^{-1}$ out to d ~300Mpc. These are detected as candidate TDFs by finding new scanning survey sources in the most recent 0.3-3months (sliding detect window) of integrated sensitivity with HET positions (all are <20") coincident with galaxy nuclei. As candidates are found, a short (1-2ksec) pointing is triggered to verify the source in the more sensitive focusing SXI telescope and to obtain a high resolution (0.15" pixels) IRT image in its 4-bands (ξ4.3). Confirmed detections then trigger spaced (~1mo.) followup imaging and spectra to follow the TDF decay and constrain the still uncertain energy release timescales. The large TDF sample built up over the 5y mission would allow detailed SXI and IRT studies of their galactic nuclei to constrain their otherwise quiescent SMBH masses from their measured bulge velocity dispersions and the M - σ relation. This then provides the "independent test" needed for the M - σ relation itself: if a TDF is found in a galaxy with a very low mass bulge, or conversely with a high-σ bulge pointing to a SMBH with M >$10^8$ $M_\odot$ (for which tidal disruption is not expected, since stars are "swallowed whole"), then the relation is called into question or is at least in need of revision.

Many other examples from Fig. 4 could be cited (e.g., outbursts of accreting stellar BHs in X-ray binaries in the Galaxy as well as Intermediat mass black holes (IMBHs) in the nearby Local Group[32], and still more classes of transients not included (e.g. "excretion" disk outbursts of Be stars with unseen BH companions producing X-ray outbursts without the characteristic X-ray pulsations seen in the Be-NS binaries thus far discovered). The era of Time Domain Astronomy with *LSST*, *SKA* and other wide-field facilities at the time *EXIST* could fly will be enhanced enormously by the unique TDA attributes of *EXIST*: rapid all-sky coverage, precise source positions and broad band imaging and spectra over the nIR to hard X-ray bands with minimal extinction.

### 2.4 Science-driven requirements for the *EXIST* mission

Given the primary science objectives (ξ2.1 – 2.3), the "flowdown" requirements for the optimum mission to *EXIST* are summarized in Figure 5:

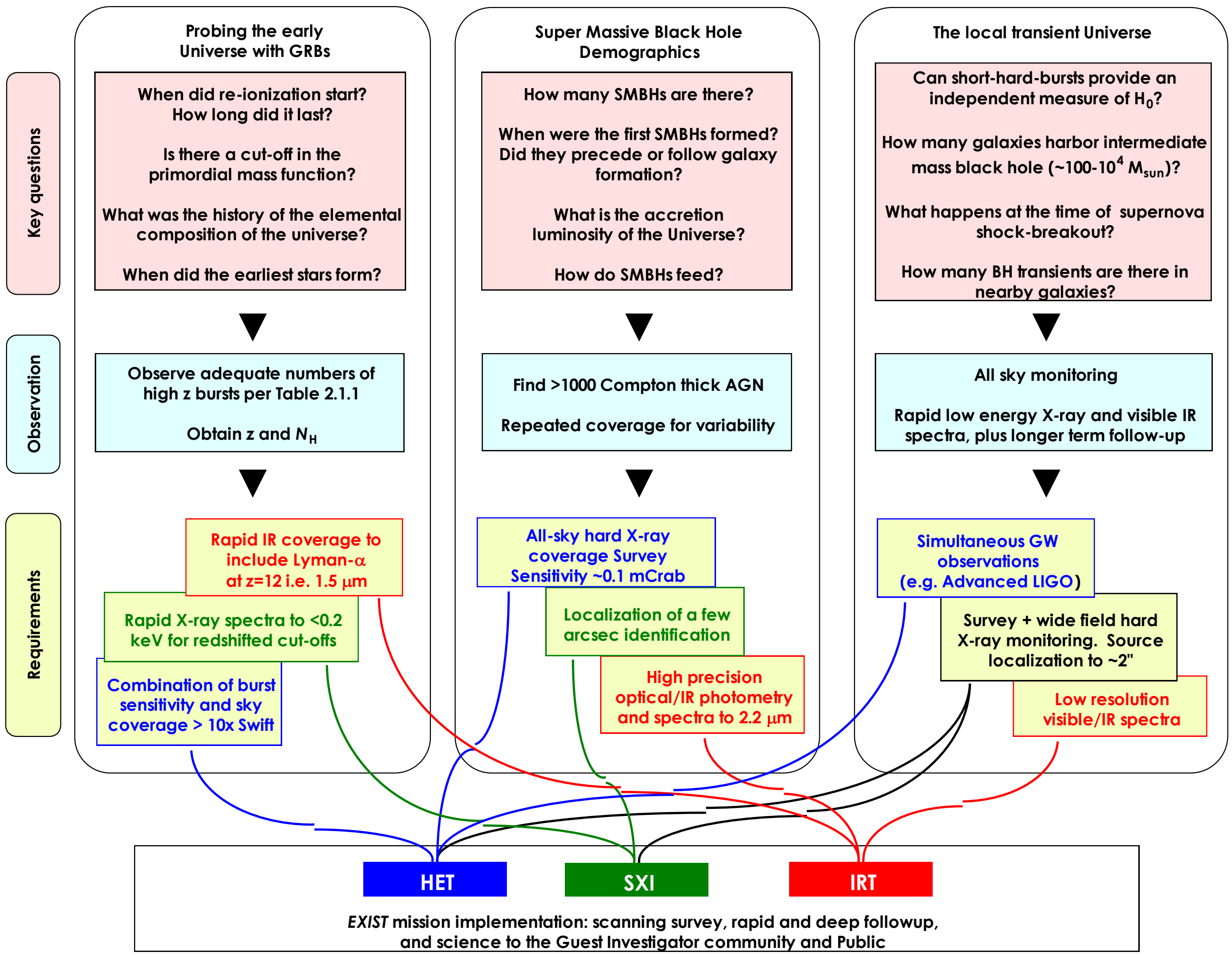

Figure 5. Flow-down from EXIST science to measurements to mission requirements.

The key questions posed by each of the three primary science objectives define a set of observations needed which in turn set the requirements for the instrumentation and mission needed. The resulting mission then has three telescopes: i) the High Energy Telescope (HET), a large area ($4.5m^2$) coded aperture wide-field 5-600 keV imager; ii) the Soft X-ray Imager (SXI), a focusing 0.3-10 keV X-ray telescope; and iii) the optical-near infra-red telescope (IRT), a high resolution imager and spectrometer covering the near-UV to near-IR (0.3 – 2.3μm) in four simultaneous bands.

## 3. OVERVIEW OF *EXIST* MISSION

The *EXIST* mission, as proposed to the 2010 Astronomy and Astrophysics Decadal Survey (Astro2010), is a Medium class mission designed for launch on an EELV launcher such as the AtlasV-401 (the minimal lift/cost launcher with 4m diameter fairing). The nominal lift capability of the AtlasV-401 and total mass (with contingency) for the mission would allow a KSC launch into a moderately low inclination ($i = 22^o$) orbit, with possible value of $i = 15^o$ depending on final launch mass. An even lower inclination ($i < 5^o$) orbit, with lower backgrounds and minimal SAA interference, could be achieved with a launch instead using an ESA Soyuz (virtually identical to the AtlasV-401 in size and capability) from the ESA Kourou launch site if ESA participation in the mission can be arranged. Given that the mission would have strong participation from Europe (particularly Italy; see ξ4.2), this desirable option would reduce NASA costs at relatively low cost to ESA. We first summarize the mission payload and then the mission plan.

### 3.1 Mission payload

The overall payload is shown in Fig. 6a, with the three primary telescopes co-aligned mounted on the spacecraft, and in Fig. 6b mounted in the launch vehicle. The top-level characteristics of the telescopes and mission are given in Table 1.

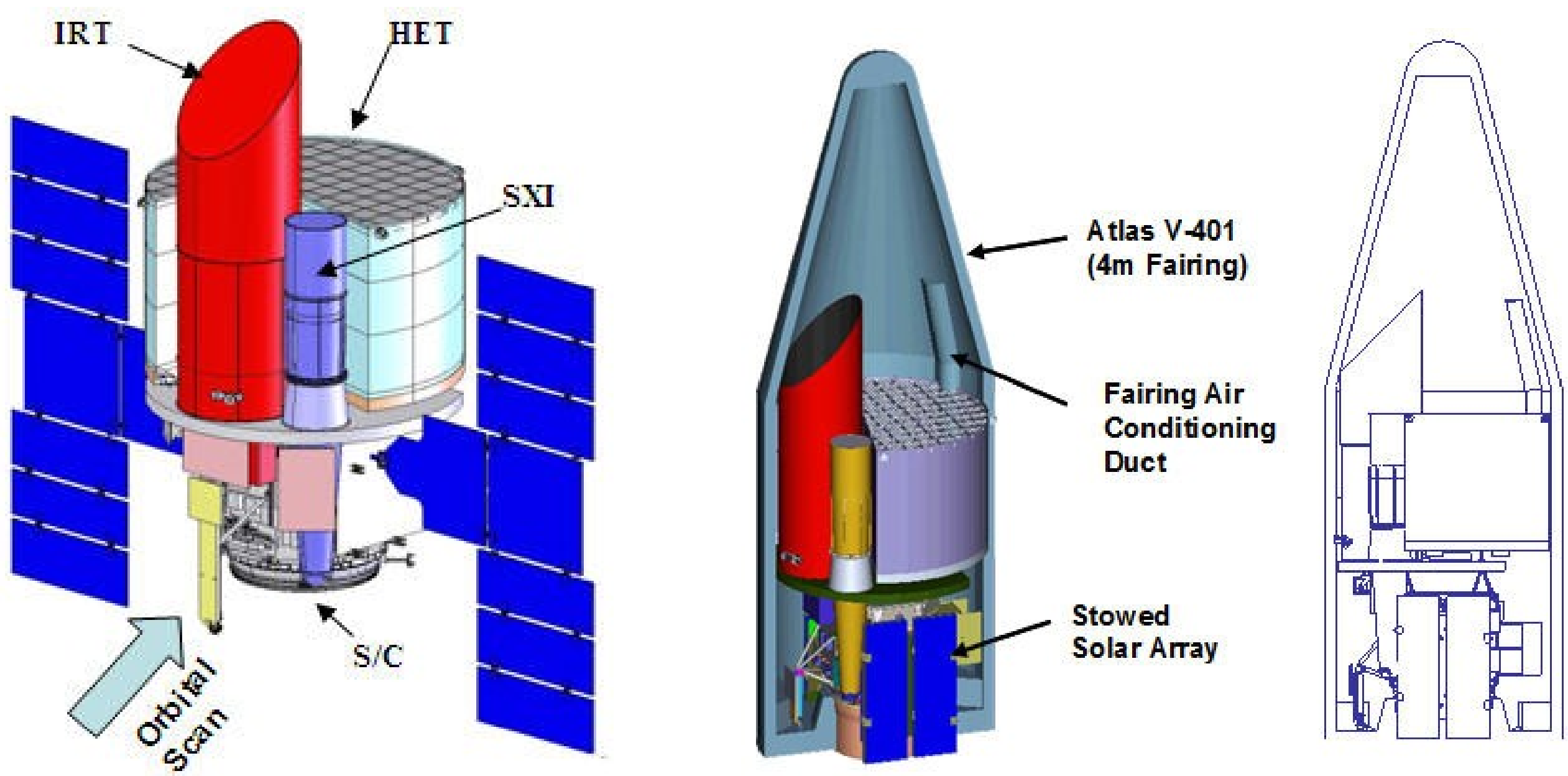

Figure 6. Left, a) *EXIST* mission and its three primary telescopes (HET, SXI, IRT; see ξ4.1, 4.2, 4.3) mounted on the spacecraft (S/C); Right, b) S/C and instruments mounted in the 4m diameter fairing of the AtlasV-401 launch vehicle.

Table 1: *EXIST* mission parameters.

| **Parameters** | **Values** |
|---|---|
| Orbit | 600 km, 15º–22º inclination, 5yr mission |
| Mode | Zenith orbital scan (2yr); inertial pointing (3yr) |
| High Energy Telescope (**HET**):<br>$4.5m^2$ Cd-Zn-Te imaging (0.6mm pixels) coded aperture | 5–600 keV, 90º×70º FoV, ≤20″ positions (90% CL)<br>0.08–0.4 mCrab (<150 keV, 1yr survey, 5σ)<br>0.5–1.5 mCrab (>150 keV, 1yr, 5σ) |
| Soft X-ray Imager (**SXI**):<br>$A_{eff}$ ~$10^3 cm^2$ Wolter-I focussing | 0.3–10 keV, 30′×30′ FoV, <2″ positions (90% CL)<br>$2\times10^{-15}$ erg $cm^{-2}$ $s^{-1}$ (10 ks) |
| Optical/IR telescope (**IRT**):<br>1.1m aperture R-C | 0.3–2.2 μm, 4′×4′, 0.3″ resolution<br>AB~24 mag in 100 sec |
| Spacecraft (S/C) | Pointing: 2″ stability; Aspect: 2″ (90% conf.) |
| Mass | 5497 kg (20% contingency) + 434 kg (propellant) |
| Power | 2803 W (30% contingency) |
| Telemetry | 65 GB/day; realtime (TDRSS) GRB downlinks |
| Launcher | EELV with 4m fairing (e.g. Atlas V-401) |
| Cost | $1.1B incl. launch +5y ($FY09, Price-H, @70% CL) |

### 3.2 Mission operation

The overall mission design is centered on the wide-field HET to detect, measure and accurately locate GRBs, SMBHs and transients in its medium-hard X-ray (5-600 keV) band. For GRBs and bright (>10mCrab) transients, the positions are computed on board (<10sec, or event duration) and if warranted (e.g., for all GRBs), a prompt S/C slew is executed if allowed by Sun avoidance (>40$^o$ offset required), Moon or Earth limb constraints. Slews are possible within ~150sec to achieve stable pointing (<2") of the HET, SXI and IRT on the <20" position computed by the HET. For GRBs or unknown transients, the SXI imaging will yield a <2arcsec position typically within 100sec, which is compared with the 100s image simultaneously acquired in 4 bands by the IRT for an obviously variable counterpart. If found, the IRT tip-tilt mirror (see ξ4.3) puts the object onto the long-slit for a spectrum (in all 4 bands) at R = 3000 if the magnitude from the 100s image are AB <20 or onto the R = 30 long slit if AB <23 or the R =30 objective prism if the ID is uncertain. If none of these conditions are met, a deeper (300s) image is obtained and the process repeated. At each stage (i.e. after first 100s pointing), SXI and IRT images (object positions and magnitudes) are sent down via the rapid TDRS link, which has already sent down the HET position for the GRB (or transient) that initiated the slew. Thus ground-based autonomous processing and comparison with deep catalogs will provide realtime checks on the onboard object recognition and can initiate a pointing correction for spectroscopy if needed (in rare cases). Details of the autonomous on board decision tree and the extent to which on board stored catalogs (down to AB ~24) will be possible, are still under study with simulations. For the GRB sample with redshifts now available from *Swift*, the above procedure would result in redshifts for all within the first ~1000sec and high-res spectra for ~70% within the first 3ksec of IRT integration.

As mentioned above, during the first 2y of the nominal 5y mission, *EXIST* is in a scanning mode, with its view axis pointed at the local zenith but ~25$^o$ North and then ~25$^o$ South of the orbital plane on alternate orbits. This allows the 90$^o$ x 70$^o$ FoV of the HET to scan the whole sky every two orbits (3h) and achieves the following advantages over a survey done with a fixed-pointing large FoV coded aperture telescope (e.g. *Swift*/BAT or *INTEGRAL*/IBIS): i) larger sky coverage per unit time, which is particularly important for long GRBs at high redshift which may have T90 durations (1 + z) times longer than their counterparts at low z; ii) longer exposure times for any given source since it is observed for ~20min (=FoV diameter/4$^o$/min scan rate) every 3h, and iii) higher sensitivity for the coded aperture imaging (see ξ4.1) since systematics in the pixel-detector plane are averaged over (much) more completely than with "dithering" of the pointing pattern as done by *INTEGRAL*. Figure 7 illustrates the orbital scan mode with the initial detection of a GRB which will trigger a slew to its position.

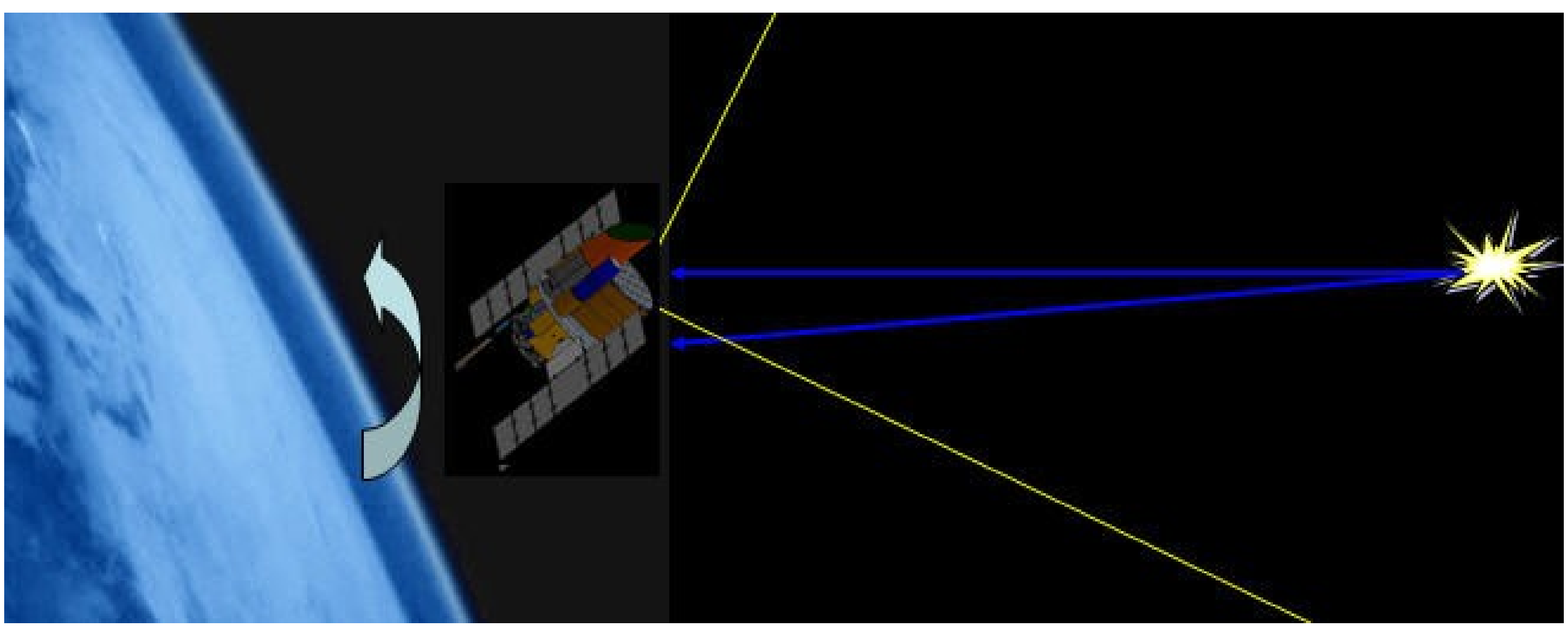

Figure 7. *EXIST* in orbital scan mode (broad arrow) detects GRB in the HET 90$^o$ x 70$^o$ FoV (yellow lines) and will slew to its computed position for pointing over the remainder of that orbit's visibility as well as following orbit(s) to measure redshift and spectrum with the IRT and low energy absorption and variability with the SXI.

Since typically ~2 GRBs per day are expected, and each is followed by ~1-3 orbits of pointing (depending on brightness) with the SXI and IRT, the scanning survey is in fact ~75% of the first 2y, with the remainder in pointing mode. During these GRB pointings, the HET can and will trigger on new GRBs, though slews to these may be inhibited for GRBs determined to be high redshift (usually within the first ~300s after trigger) so that their high resolution nIR spectra can be obtained without interruption. Due to its small (4arcmin; Table 1) FoV and readout electronics based on *JWST*/NIRSPEC (see ξ4.3), the IRT only operates in pointed mode and so is taking data for ~25% of the scanning mission phase and 100% of the following 3y pointed mission phase, and thus for ~3.5y of the 5y mission. The SXI imaging spectrometer is, like the HET, photon counting and so both are active for 100% of the mission. The SXI is then conducting its own sky survey during the scanning mission, with its 0.5$^o$ FoV covering approximately half the sky (covering Declination range δ ~ ±45$^o$) every 6 months (see ξ4.2).

For the 3y pointed phase of the mission, the primary targets are i) the candidate Blazars, particularly those with high z, ii) the candidate obscured and Compton thick (CT) AGN and iii) the candidate TDFs and other extreme variables found in the HET-SXI scanning survey. The blazars are found by selecting on source colors for sources with hard spectral and $\upsilon F_\upsilon$ fluxes in HET (e.g. 10 - 20 vs. 20 – 40 keV) that exceed those in SXI (e.g. 1.5 – 3 vs. 3 – 6keV), which would point to SEDs similar to those for the BAT blazars as shown by G10. Obscured and candidate CT AGN are similarly identified from the SXI survey as sources with low energy absorption. Any sources detected only by HET (at 5-10 keV and above) and not by SXI are candidates to be either Blazars or CT AGN and so are also prime targets for the pointed followup survey. The total expected AGN candidates (i and ii) are ~19,000 and 1400, respectively (Fig. 3) although the number of obscured and CT AGN are expected to be enhanced significantly from the results of the SXI survey. For identification and classification by the SXI and IRT of the sources, the pointing survey will typically require 1-2 orbit pointings (2ksec each). Thus with 16 orbits/day and so ~20 objects/day observed (2 objects can be observed each for ~2ksec on each orbit), ~15000 objects can be classified and studied in a 2y pointing program. This will enable the key goals outlines in ξ2.1- 2.3 and summarized in Fig. 5 to be accomplished in ~2y. Since the remarkable capabilities of the mission will almost certainly lead to discoveries and demands for followup, it is essential this be done in year 3 (and beyond) of the pointing survey. Margin is also needed for the innovative proposals for combined (or separate) scanning vs. pointing "medium" and "deep" surveys and followup that will be part of the Guest Investigator (GI) program proposed for *EXIST* (ξ6) to be accomplished in the nominal 5y mission – which, with no rapid consumables, could be extended indefinitely.

## 4. THE TELESCOPES TO *EXIST*

The three major instruments/telescopes proposed for the *EXIST* mission are summarized in Table 1. Here we provide brief summaries, with a figure and Table for each; details are given in the references cited.

### 4.1 High Energy Telescope (HET)

The current design for the imaging CZT and the large area coded aperture telescope, the HET (see Fig. 1 and Table 2), has been described by Hong et al[34,35,36]. Here we provide an overview including some aspects not previously discussed.

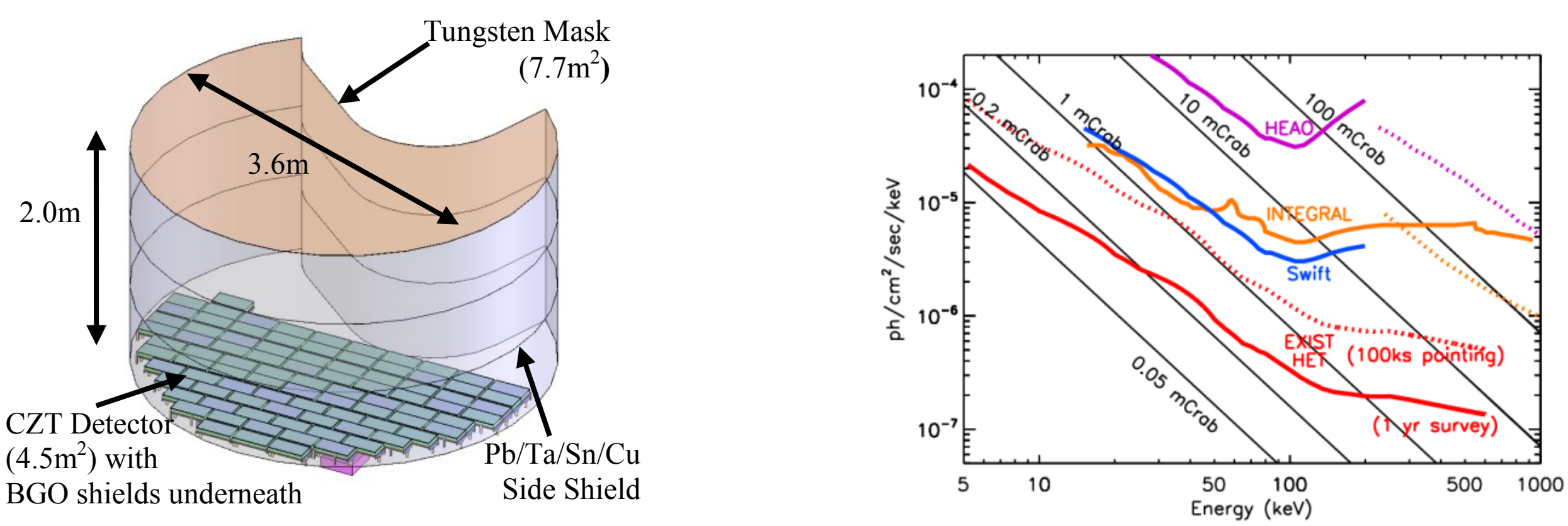

Figure 8. Left, a) Overview of HET, showing principal components and imaging CZT array of 88 DMs (see text); and Right, b) HET sensitivity for 1yr scanning survey vs. corresponding all sky survey sensitivities for *Swift*/BAT, *INTEGRAL*/IBIS and HEAO-A3, each scaled for FoV and sensitivity if the sky is uniformly covered. The HET sensitivity for the 2yr scanning survey is a factor √2 better (lower on plot), and the 3y pointing survey (which covers full sky) is a factor of ~2.5 below the 1yr scanning survey. The 100ksec pointing sensitivity scales as $\sim T_{exp}^{-0.5}$ for exposure times $T_{exp}$.

We comment first on the fine pixel readout for the HET and its scale: the ~12.5M spectroscopic pixels represented by 4.5m$^2$ of imaging Cd-Zn-Te (CZT) detector crystals (each 2 x 2 x 0.5cm and with 32 x 32 anode pixels readout by a direct-bonded ASIC on each) are *not* a reliability or calibration challenge. The HET is, instead, a *highly redundant* detector, each crystal imaging basically (semi - ) independently of the others, as do the ~32,000 separate CZT crystals (each non-pixelated) on the BAT detector on *Swift*. If ~10% of the CZT crystals (or their ASICs) are "dead", this only represents a ~5% loss of sensitivity for this (or any) coded aperture imager. The 0.6mm detector pixels read out the 1.25mm mask pixels (<100keV) and coarse mask pixels (13.85mm) at high energies (>150 keV) of the hybrid coded mask (see Hong et al 2009). Calibration is measured by a tagged $^{241}$Am source, as for BAT, but now also with independent calibration by an internal pulser on each pixel as well as the advantages of scanning. And, with an active rear shield (7mm thick BGO behind each 16 x 32cm detector module (DM; see Fig. 8), together with close-tiling (see Fig. 12b of Hong et al[36]) of the CZT that avoids higher background on the crystal side-walls, as experienced by BAT, the HET would have significantly lower backgrounds (particularly at >40 keV) due to Earth albedo as well as capability to measure high energy spectra (to ~3 MeV) and $E_{peak}$ for GRBs. Finally, the multi-pixel readout (hit pixel plus near-neighbors plus reference pixel), combined with ~0.7mm depth sensing for each event at >200 keV (Allen et al[37]), means that Compton imaging and polarization measurement is possible, particularly for GRBs.

Table 2. HET parameters.

| Parameters | Values |
|---|---|
| Telescope (coded-aperture) | 4.5m$^2$ CZT (0.6mm pix, 11.5Mpix)<br>7.7m$^2$ tungsten mask (1.2/14mm pixels) |
| Energy Range | 5 – 600 keV (imaging CZT)<br>200 – 2000 keV (BGO for GRBs) |
| Sensitivity (5σ) (~1y survey) | 0.08– 0.4mCrab (<150keV)<br>0.5–1.5mCrab (>200keV) |
| (10s on-axis) | ~24mCrab (<150keV) |
| Field of View | 90$^o$ × 70$^o$ (out to 10% coding) |
| Angular Res. Centroiding | 2.4′ resolution<br><20″ for >5σ source (90% conf. rad.) |
| Sky Coverage | Full sky every two orbits |
| Spectral Res. | 3 keV<br>(3% at 60 keV, 0.5% at 511 keV) |
| Time Res. | 10 μsec |
| Heritage | *Swift*/BAT, *INTEGRAL*/IBIS ( and *Fermi*/LAT (for orbital ops) |

The successful balloon flight of *ProtoEXIST1*, the first-generation prototype for the imaging CZT, with 2.5mm pixels, read out by an ASIC below and close-tiled to form a reduced size DM, is discussed by Hong et al[36] and the detector and telescope integration by Allen et al[38]. The currently ongoing development of the close-tiled CZT imager with 0.6mm pixel size for *ProtoEXIST2* and then the near-final prototype, *ProtoEXIST3*, are also described in Hong et al[36] 2010a, with a balloon flight test of *P2-P3* planned for Spring 2012.

### 4.2 Soft X-ray Imager (SXI)

The overall design for the SXI is given by Taliaferri et al[39], with details for its grazing incidence focusing Wolter-I telescope (Fig. 9, top) given by Basso et al.[40], and for the SXI camera by Uslenghi et al.[41] (Fig. 9, bottom right).

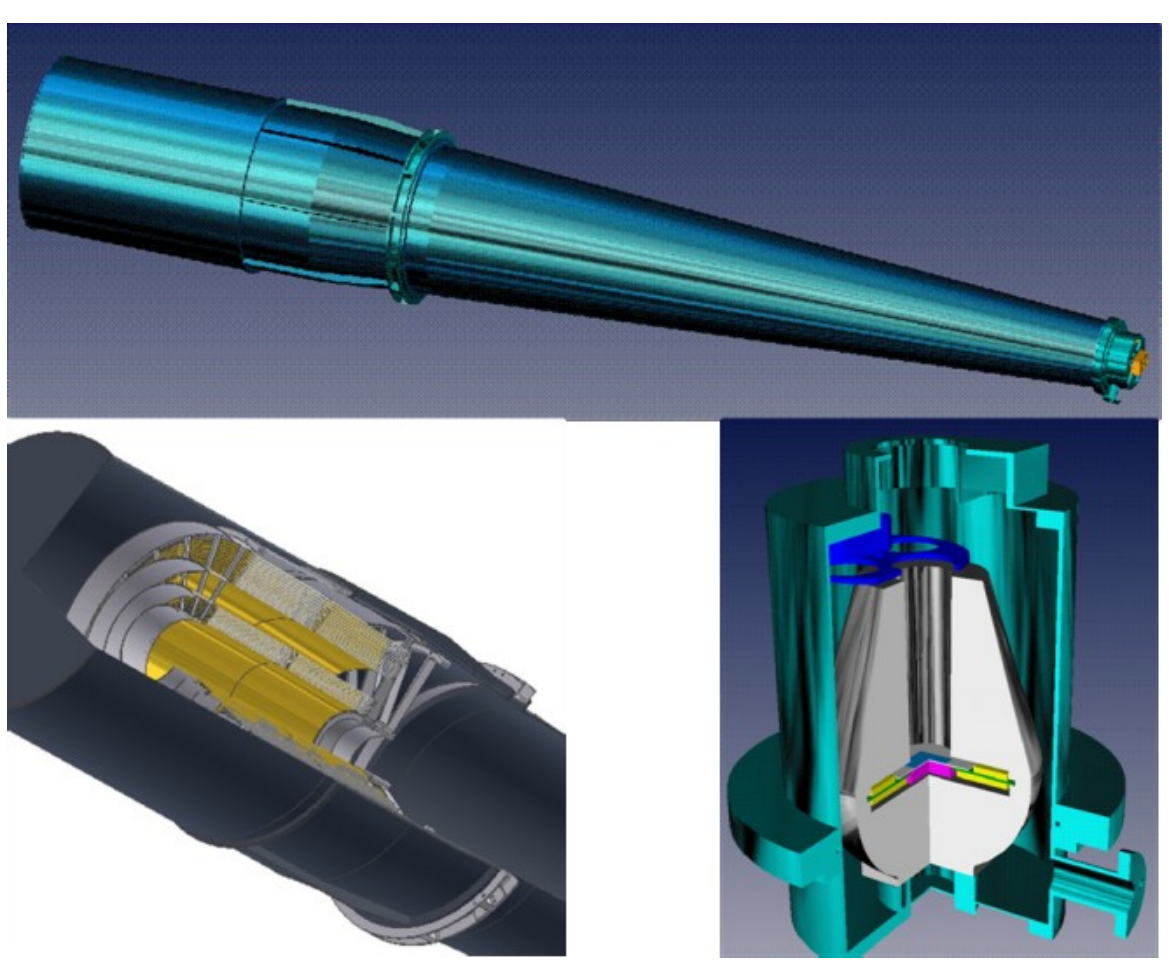

Table 3. SXI parameters.

| Parameter | Baseline |
|---|---|
| Mirror | 26 shells |
| Angular Res. | 20″ @ 1 keV |
| Energy range | 0.1 – 10 keV |
| Dia. of mirrors | 60 cm |
| Focal length | 3.5m |
| Detector type | PN-type CCD (or DEPFET) |
| FoV, Detector | 30×30 arcmin$^2$, 3×3 cm$^2$ |
| Energy Res. | E/ΔE = 47 at 6 keV |
| Readout speed | 5 – 10 ms (or 1 ms, if DEPFET) |
| Instrument effective area | 950 cm$^2$ at 1.5 keV,<br>>100 cm$^2$ at 8 keV |
| Sensitivity (10$^4$s) | 2.0×10$^{-15}$ erg cm$^{-2}$ s$^{-1}$ |

Figure 9. SXI telescope (top, lower left) and camera (lower right). The SXI design and fabrication would be led by Italy.

The effective area of the SXI and overall sensitivity (Table 3) is approximately that of one PN telescope on XMM-Newton, so that it is approximately 10X more sensitive than the XRT on *Swift*. The focal plane camera design is a 384 x 384 pixel array with 75/80 μm pixels depending on whether it is a CCD (e2V) or DEPFET (MPE) detector as planned for eROSITA and possibly IXO (Treis et al[42]). Either camera could be operated at -50C with only passive cooling though a thermo-electric cooler would be provided to minimize solar constraints. Both are fast enough to allow event-based readout as the standard mode, but the DEPFET, with still higher count rate limits, would allow greater dynamic range for the early afterglows (or even late prompt) emission of GRBs. Both camera designs are fast enough to enable the SXI to remain fully operative during the 2y scanning mode survey. This allows the SXI to conduct a sensitive half sky survey during the 2y scanning mission. The sky coverage and approximate sensitivity is shown in Fig. 3 of Natalucci et al[43] and gives half-sky coverage every 6mo. over the Declination range ±40-45$^o$ for an assumed orbital inclination of i = 20$^o$ and offset of ±25$^o$ on alternative orbits. Each source in this half-sky survey is exposed for ≥200s (those near the Dec limits get more exposure), giving a survey flux limit of F(0.3-10keV) ≤5 x $10^{-14}$ erg $cm^{-2}$ $s^{-1}$, or a factor of ~15 below that for the corresponding HET survey for an assumed power law spectrum with photon index 2. Thus essentially all HET survey sources (within the half-sky covered by SXI) are not only detected, but have their spectra <10 keV measured or well constrained by the SXI. This enables the selection of optimum Blazar and CT-AGN candidates for the targeted HET-SXI-IRT followup survey (ξ3.2) as well as identification and study of all classes of sources found in the survey. Each SXI survey source (in the HET sample) is located to <2arcsec positional accuracy, making the optical-nIR identification and detailed spectral studies with the IRT possible from the same pointing survey.

### 4.3 Optical – near Infra Red Telescope (IRT)

The IRT is the truly novel telescope on *EXIST*. It combines unprecedented near-IR sensitivity, obtained by passively cooling (on dark sky) the primary mirror and subsequent mirrors to -30C, which reduces thermal backgrounds at 2μm by a factor of ~$10^3$, thereby achieving zodiacal light background levels that give sensitivities and speeds at 2μm that are 10X that of the 10m Keck telescope. The IRT design by Kutyrev et al[44] has an innovative combined imaging and spectroscopy (low and high resolution) focal plane for simultaneous imaging or spectroscopy in 4 bands from 0.3 – 2.2μm (see Table 4) with no moving parts or filter choices to be made. The telescope itself is a 1.1m telescope designed by ITT for the *GeoEye* mission now in orbit (Fig. 10a) but modified (for IRT) as a simpler Ritchey-Chretien. The focal plane design (Fig. 10b) and overall imaging vs. spectroscopy optics (Fig. 11a) and field of view allocations (Fig. 11b) were designed at GSFC by Kutyrev et al[44]. The focal plane enclosure is actively cooled ~15K below the passively cooled ambient mirror temperature (240K). The two IR detectors are H2RG Hg-Cd-Te detectors (2K x 2K) as developed for NIRSPEC and NIRCAM on *JWST* and the two visible HyVISI detectors employ active pixel Si detectors and the same Teledyne SideCar ASIC readout as the H2RG detectors. Thus all focal plane elements have high TRL. The tip-tilt mirror and 10Hz servo system (Fig. 11a) provides autoguiding (0.15" stability) over a 6" range so that the <2" inertial pointing stability of the S/C is readily accommodated for precision pointing. The tip-tilt is also used for small offsets – e.g. from a ~1-2arcsec SXI source position to center up an object on a 0.3" slit.

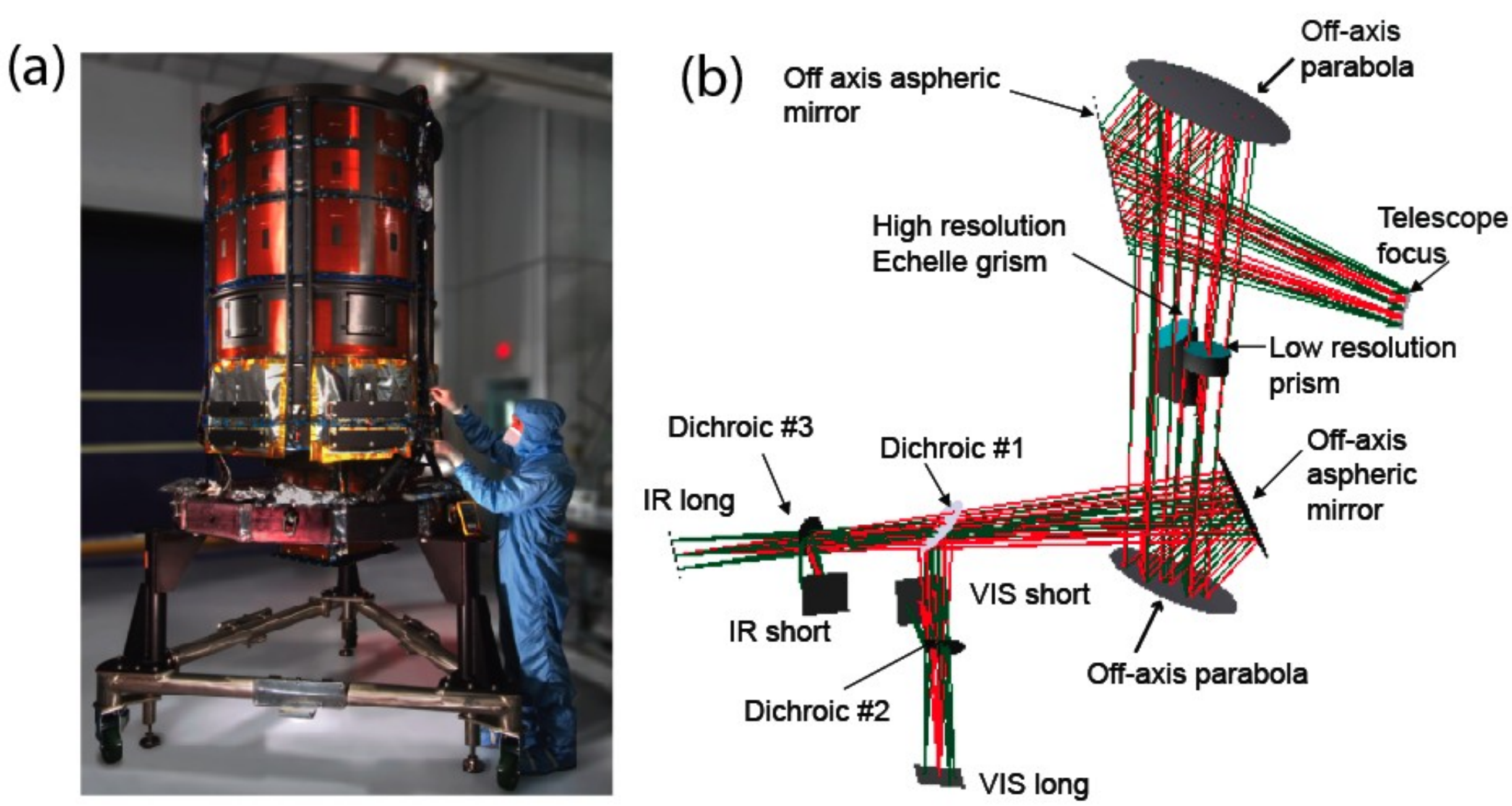

Figure 10. Left, a) 1.1m optical Earth-imaging GeoEye telescope and IRT prototype; Right, b) optical-nIR focal plane design for IRT (from Kutyrev et al[44]).

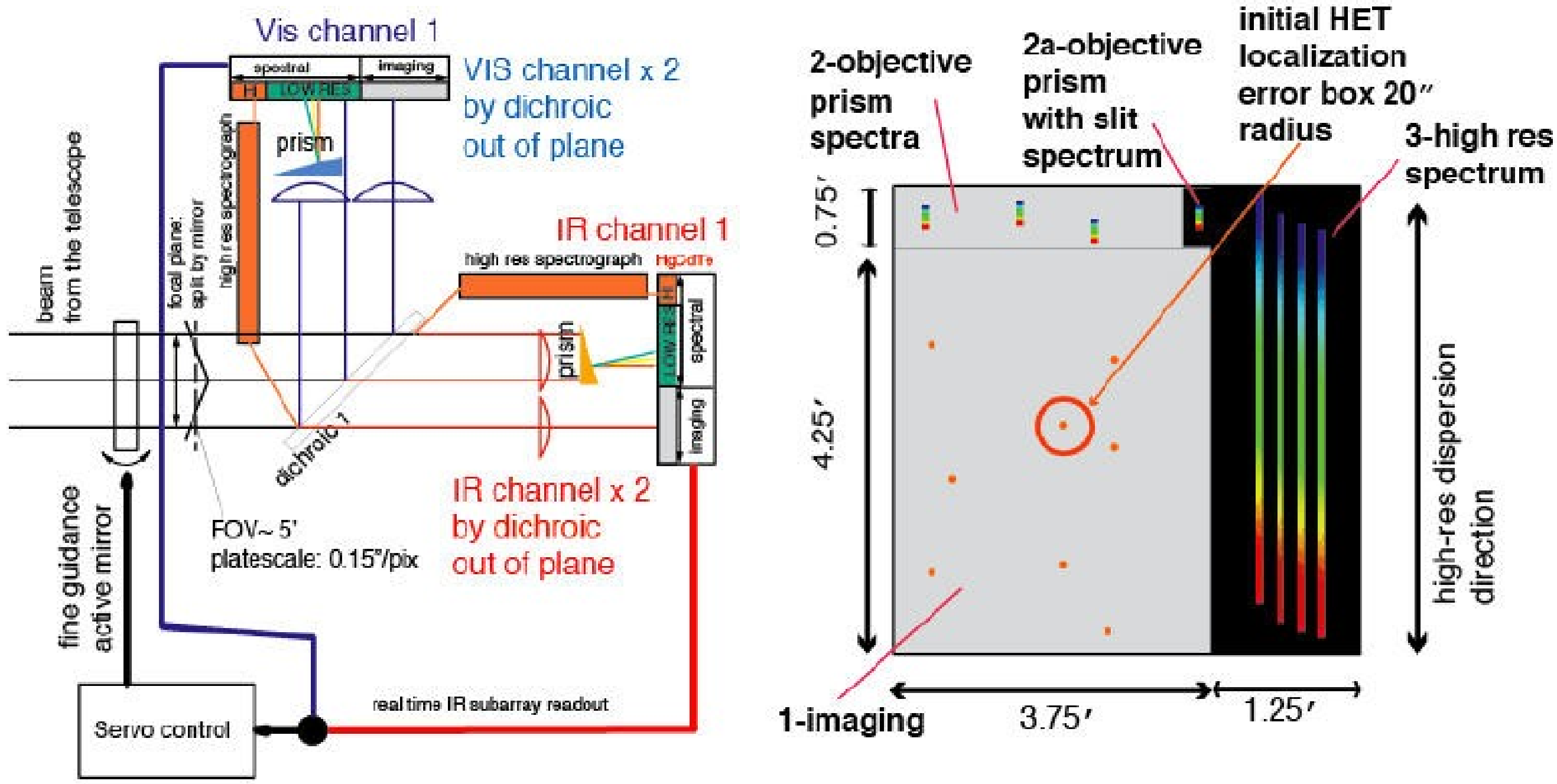

Figure 11. Left, a) Optical and IR (2 bands each) simultaneous imaging and/or spectroscopy with tip-tilt autoguiding. Right, b) Allocation of 2K x 2K detector area (optical and IR) for imaging vs. spectroscopy (from Kutyrev et al[44]).

Given this optical design and the IRT parameters summarized in Table 4, the nominal sequence of IRT observations for a GRB, following a rapid slew to a <20" source position computed by the HET, would then be as shown in Table 5.

Table 4. IRT parameters.

| **Telescope** | 1.1m aperture R-C Cassegrain | | |
|---|---|---|---|
| **Ang. Res.** | ≤0.3″ PSF (0.15″/pixel pl. scale) | | |
| **Spectral Bands (4)** | 0.3 – 0.52, 0.52 – 0.9μm (*HyViSi*)<br>0.9 – 1.38, 1.38 – 2.2μm (*H2RG*) | | |
| **Mode** | **FoV** | **Spec. Res.** | **AB @S/N≥5σ (int. time)** |
| Imaging | 3.75′×4.25′ (16arcmin$^2$) | ~3 | 24 (100 sec) |
| Low Res. Obj. Prism | 3.75′×0.75′ (2.8arcmin$^2$) | ~30 | 22 (300 sec) |
| Low Res. Single Slit | 20″ long slit | ~30 | 23 (300 sec) |
| High Res. Single Slit | 4″ long slit | 3000 | 19 (2 ksec) |
| | | | |

Table 5. Autonomous IRT measurement of GRBs

| **Seq. #** | **Time** | **Observation** |
|---|---|---|
| 1 | 100s | Initial 4-band image (AB~24) |
| 2 | 300 | Initial R=3000 spectrum if 1 AB<18 obj. @<2" from SXI |
| or 2a | 300 | Initial R=30 slit spectrum if 1 AB<22 obj. @<2" from SXI |
| or 2b | 300 | Initial R=30 objprism spec if 1 AB<22 obj. @<2" from SXI |
| 3 | 3ksec + | R=3000 slit spectra if AB <19 |

The IRT observation sequence for a GRB identification and followup spectroscopy shown in Table 5 is for a "clean" high latitude GRB field, which will be typical for most. These should be possible to do autonomously, though with object catalogs (positions and approximate magnitudes) sent down via TDRS, some cases will require command uploads to do pointing offsets onto the likely ID. However for crowded or obscured fields (e.g. nearer the galactic plane, or dense clusters), the optical or nIR counterpart will not always be obvious either objective prism or several repeated imaging exposures may be required to identify the counterpart before committing to the long slit spectrum desired. The observation sequence shown extends only through the first orbit. Most GRBs will be followed for at least 2 orbits; all at z >4 -5 will be followed for ≥3 orbits. In the second orbit, and subsequent orbits (if obtained later), a deeper set of images (4 bands) would also be obtained for host galaxy idenfitifcation and morpholoty or limits (e.g. 4 orbit exposures should reach AB ~27) to guide followup *JWST* observations (for z >6-8 GRBs).

For SMBH/AGN and transient source IRT observations, the "typical" observation sequence is similar to that above, though here the final identification may require several trial spectra (e.g. for high-z Blazars, which will in many cases be continuum spectra objects and so not readily distinguished from metal poor stars). However for most newly discovered AGN in the scanning survey (Fig. 3), the typical range of X-ray/optical flux values would predict optical/nIR magnitudes AB ~19-20 so that their initial identification as emission line objects by using the R = 30 objective prism or slit (depending on positional uncertainties, particularly for the half-sky without SXI survey coverage) will provide rapid (<300s) identifications that then allow (for most) R =3000 spectra for detailed studies (e.g. reverberation line mapping).

## 5. SPACECRAFT AND MISSION PARAMETERS

As part of the ASMC (Astrophysics Strategic Mission Concept) Study carried out for *EXIST* in 2008, during which the detailed design of the three primary telescopes and their instruments was developed with an engineering study at the NASA/GSFC Instrument Design Lab (IDL). The spacecraft (S/C) and systems resources required to support (power, point, deliver data, etc.) the telescopes were also studied in the Mission Design Lab (MDL). The principal components of the S/C and instruments are shown in Fig. 12, and the S/C and mission summary is given in Table 6.

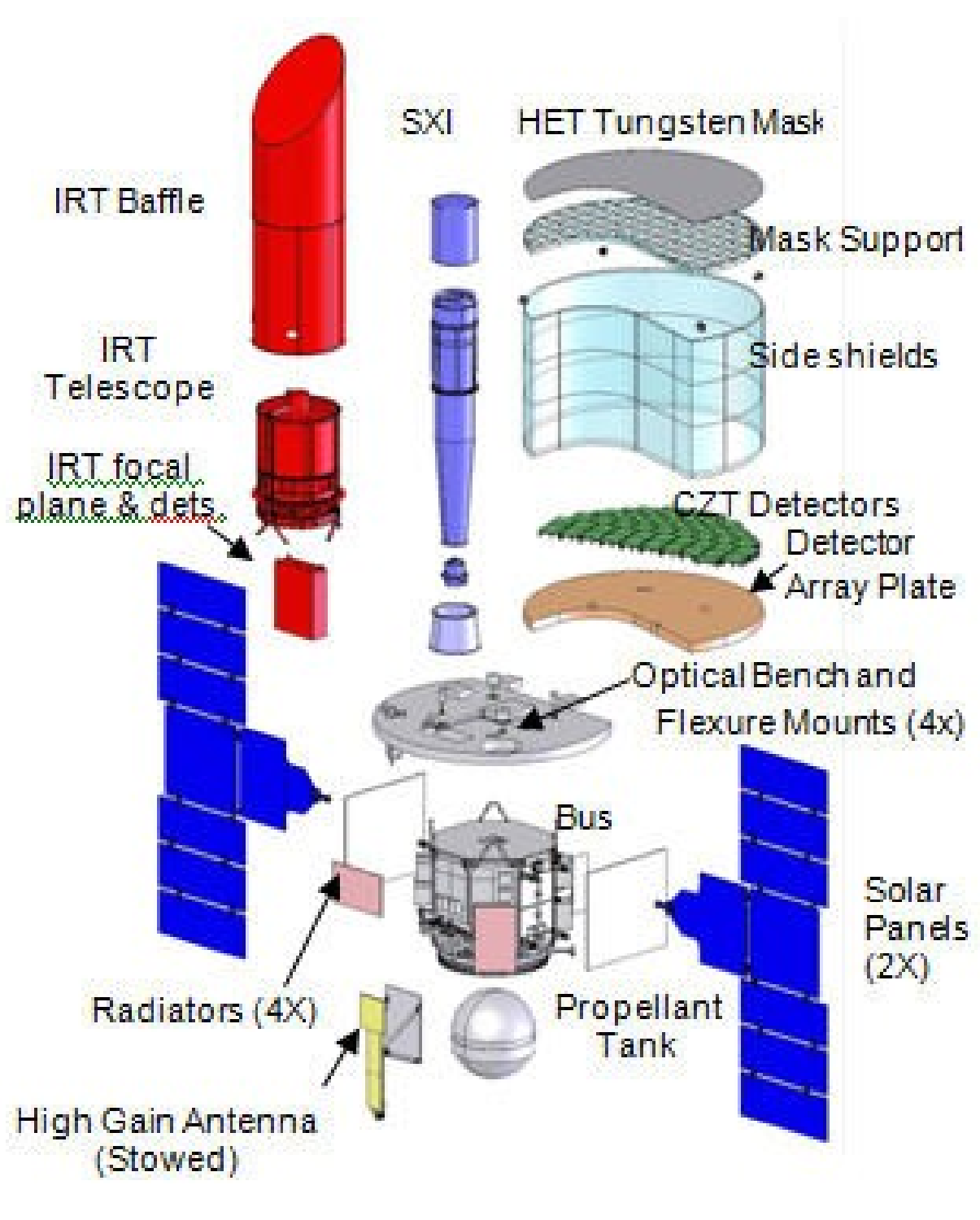

Table 6. Spacecraft and mission parameters.

| Parameter | Req't | Capability | Margin |
|---|---|---|---|
| Orbit | 600 km, ≤ 22º | Comply | N/A |
| Launch Mass Atlas V-401 | 5932kg. CBE+ Contingency | 8000 kg @ 22º | 35% |
| Ptg. Control<br>Knowledge<br>Stability | 10″<br>5″ (3σ)<br>1-2″ | 5″<br>< 3″ (3σ)<br>< 1″ | 100%<br>67%<br>TBD |
| Slew Time | 45º in 180 s | Comply | N/A |
| Avg. Power | 2803 W | 3645 W | 30% |
| Data Storage | 130 Gbits | 300 Gbits | 130% |
| Science Downlink | 200 Mbps | Comply, TDRS Ku-Band | 3.2 dB |
| Commanding Uplink | Not defined | 2kbps TDRS SSA, GN; TDRS MA | 3.0, 32.0, 3.7 dB |
| SOH Telemetry Downlink | Not defined | w/ science or 4 kbps/1kbps TDRS SSA/MA; 1 Mbps GN | 2.5, 0.1, 14.7 dB |
| Propulsion | Controlled de-orbit, 160 m/s | 202 m/s with full tank | 26% |
| Lifetime | 5-yr | Comply, full redundancy | N/A |

Figure 12. Expanded view of major components of spacecraft and the three telescopes and their integration. The IRT baffle is both a light shield and thermal control system, routing HET heat to S/C radiators to maintain IRT mirrors at 240K.

# 6. MISSION OPERATIONS, DATA PROCESSING AND GUEST INVESTIGATIONS

## 6.1 Mission operations

The data command and data flow routing to the *EXIST* mission and both the Mission Operations Center (MOC), at GSFC, and the Science Operations Center (SOC), at Harvard/SAO at the Harvard-Smithsonian Center for Astrophysics, are shown in Fig. 13. Given the high data rate from the HET and the scanning requirement for high time resolution, photon event data are brought down rather than binned detector data. Real-time data for GRBs are brought down for rapid distribution (within ~10sec of a burst) via the TDRS-S band and include prompt GRB positions (RA, Dec), spectra, lightcurves as well as detected source positions and fluxes/magnitudes for the HET and SXI immediately after trigger, and for all three telescopes for the autonomous series of imaging and spectra (e.g. Table 5) which immediately follow. These rapid release SXI and IRT data for GRBs and bright transiens will culiminate with the redshift or compressed spectra so that particularly for high-z bursts, very large telescopes on the ground (e.g. GMT, TMT and eventually ELT) can obtain additional deep nIR spectra. GRBs will be detected at ~1-5X per day, and their full data (along with all data from both the scaning and pointing modes) will be brought down via the TDRS-Ku band link ~4-5X per day. In addition to the full data for the rapid alerts for GRBs and bright transients, the full sky data will be brought down via these same TDRS-Ku passes, so that full sky data (during scan mode) will be available typically within ~6h of receipt.

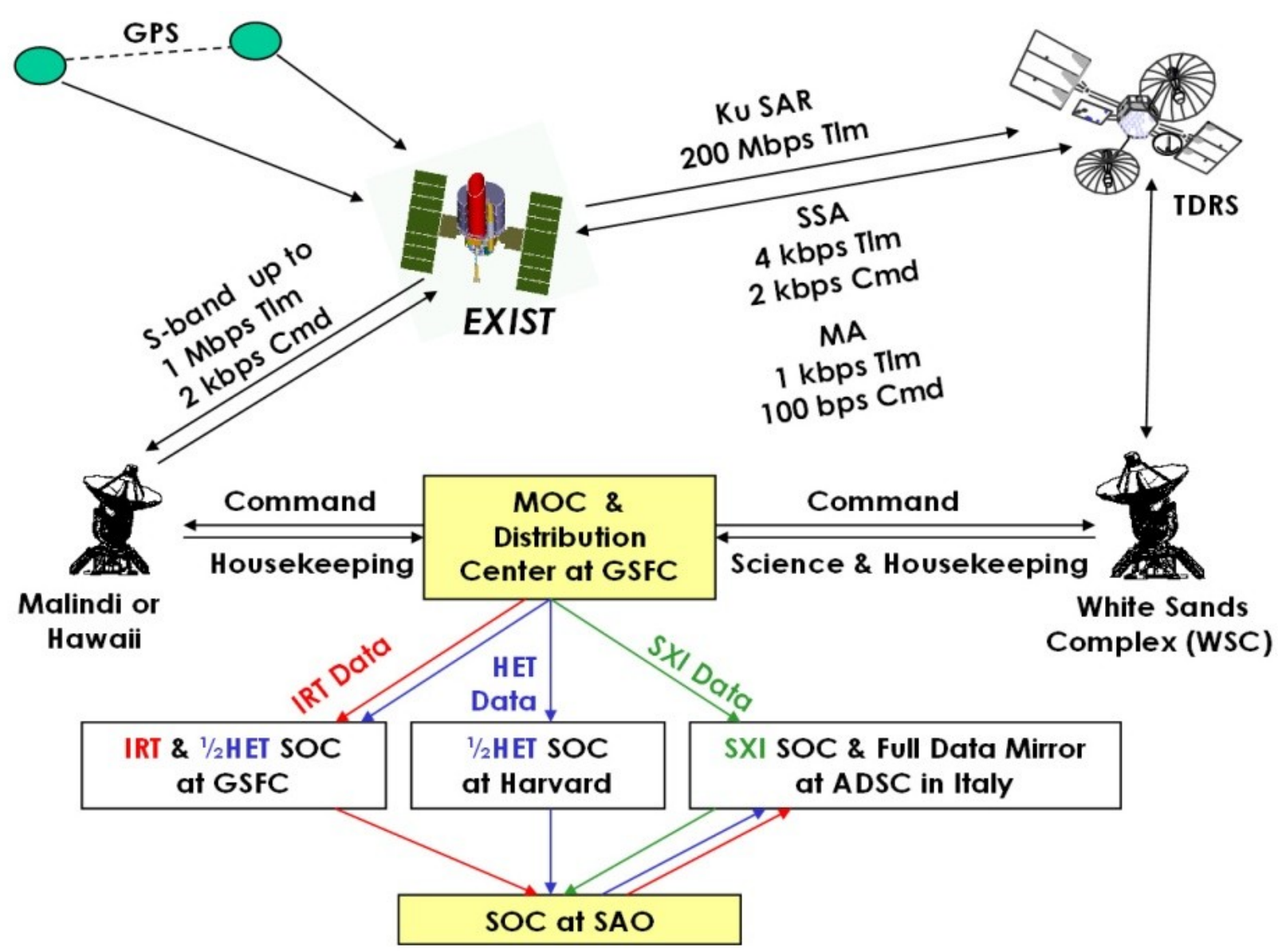

Figure 13. Data flow and commanding links for *EXIST*. GRB data (positions, images, spectra) are brought down in real time via TDRS-S band link. Ground stations (e.g. Malindi or Hawaii) are used for backup and additional capacity as needed; the full data throughout the mission are brought down via TDRS-Ku band links (to White Sands) ~4-5X per day.

Commands are all sent from the MOC and are relatively infrequent: during the scanning survey (only if small offsets are needed to correct autonomous source identification in crowded fields). Command loads during the pointing survey and followup mission phases are prepared at the SOC and issued by the MOC.

The mission will not only generate large numbers of transient alerts (using VONet protocols) for the community but will respond to transients of particular interest discovered by LSST, SKA, Advanced LIGO and other wide-field facilities operating in the same time-frame. *EXIST* will have sensitivity and utility for TDA generally, and programs for responding to ToOs will be a major part of the Guest Investigator (GI) program.

During the nominal 3y pointed mission phase, and followup of scanning survey discoveries (ξ2.2), it is likely that most of the observing program will be competed in the open GI program.

### 6.2 Data processing

The continual large volume of data from HET, particularly during the scanning mission phase, require computational resources which are already in development by the *Swift*/BAT Slew Survey (BATSS) program[45]. Data processing for each of the science telescopes will be led by the instrument team groups that have built the instruments (e.g. Harvard/GSFC for HET; INAF-Italy/MSFC for SXI; and GSFC for IRT). Initial telemetry processing (Level 0) will be done at GSFC and Level 1 (calibrated) data processing performed at the SOC (where all calibration data will be maintained). Final Level 2 (images and spectra) data processing and distribution to the community are done at the SOC.

### 6.3 Guest Investigator program

A large GI program is anticipated for the *EXIST* mission and has been included in cost estimates for the total mission. Although most of the science discussions in ξ2 focused on survey science, it is likely that a significant fraction of the GI program will concentrate on detailed studies of individual or (relatively small) groups of objects to answer or test the very questions that the survey(s) have identified. Likewise, with the powerful multiwavelength capabilities of *EXIST*, it is likely that considerable interest will be in the unique multiwavelength broad band coverage possible. Finally as TDA expands even more to a major sub-field of astronomy, it is likely that timing or generalized variability studies will be significant.

The actual types of GI programs will be similar to those now familiar from *Chandra* or *XMM-Newton*, for deep studies of individual objects or limited fields; and *Swift*, for multi-wavelength and rapid followup coverage of transients or broader studies of classes of objects; and *Fermi*/LAT, for its all sky coverage and scanning survey mode of both dividing the sky (e.g. specific fields) or broad classes of objects (e.g. Blazars).

With the SXI provided by the Italian Space Agency, ASI (as proposed), and with significant contributions to the mission that have been discussed (very preliminary) with other potential partners, *EXIST is very much an international mission and will be open to all.*

## 7. SUMMARY

The *EXIST* mission offers a uniquely powerful, multiwavelength and multi-scale (both spatial and temporal), new resource for unlocking some of the most fundamental astrophysical problems. Its in situ measurements of the Early Universe by prompt detection, imaging and spectra of GRBs – from hard to soft X-rays to the near Infra-red – will enable the most direct probes and measurements of the first (or nearly first) stars and galaxies. That objective alone should justify such a mission, but it also provides the tool needed to study the most extreme black holes – from the first supermassive which might be revealed as luminous blazars, to those that are obscured at lower X-ray energies, to those that are dormant and may lurk in (bulgeless) galaxies that by our current paradigm should not contain them. And it would open the high energy transient Universe as never before studied.

After a decade of intensive study, and successive improvements to maximize its science return for minimal cost, the mission is well understood. It is ready to go. It is time to *EXIST*.

## ACKNOWLEDGEMENTS

The Study for *EXIST* was supported by NASA-ASMC grant NNX08AK84G, NASA-APRA grant NNX09AD76G and ASI (Italy) grant I/088/06/0. We thank many others on the *EXIST* team who contributed to the science and technical case, and in particular Dom Conte (General Dynamics) for his leadership on the spacecraft and systems design and Craig Golisano (ITT) for his work on the IRT.